\documentclass[aps,prl,twocolumn]{revtex4}
\usepackage{amssymb}

\usepackage{amsmath}
\usepackage{graphicx}

\newcommand{\eq}{\begin{equation}}
\newcommand{\fine}{\end{equation}}

\begin{document}

\title{Experimental sub-Rayleigh resolution by an unseeded high-gain optical
parametric amplifier for quantum lithography.}
\author{Fabio Sciarrino$^{2,1}$, Chiara Vitelli$^{1}$, Francesco De Martini$^{1}$,
Ryan Glasser$^{3}$, Hugo Cable$^{3}$, and Jonathan P. Dowling$^{3}$ \\
$^{1}$Dipartimento di Fisica dell'Universit\'{a} ''La Sapienza'' and
Consorzio Nazionale Interuniversitario per le Scienze Fisiche della Materia,
Roma 00185, Italy\\
$^{2}$Centro di\ Studi e Ricerche ''Enrico Fermi'', Via Panisperna
89/A,Compendio del Viminale, Roma 00184, Italy\\
$^{3}$Hearne Institute for Theoretical Physics, Lousiana State University,
Baton Rouge, LA70803}

\begin{abstract}
Quantum lithography proposes to adopt entangled quantum states in order to
increase resolution in interferometry. In the present paper we
experimentally demonstrate that the output of a high-gain optical parametric
amplifier can be intense yet exhibits quantum features, namely, sub-Rayleigh
fringes, as proposed by\ Agarwal et al. (Phys.\ Rev. Lett. \textbf{86}, 1389
(2001)). We investigate multiphoton states generated by a high-gain optical
parametric amplifier operating with a quantum vacuum input for a gain values
up to 2.5. The visibility has then been increased by means of three-photon
absorption. The present article opens interesting perspectives for the
implementation of such an advanced interferometrical setup.

\end{abstract}

\maketitle

\section{Introduction}

Since the early days of Quantum Electronics, non-linear optics has
played a basic role both for its relevance as a fundamental
chapter of modern science and for its technological applications
\cite{DeMa05}. Nonlinear parametric processes, due to the peculiar
correlation properties of the generated photons, have been adopted
to investigate the quantum properties of electromagnetic fields.
In the last few years it has been proposed to exploit entangled
quantum states in order to increase the resolution in quantum
interferometry, specifically, for quantum lithography
\cite{Boto00} and to achieve Heisenberg limited resolution
\cite{Holl93}. In such framework, particular attention has been
devoted to the generation of NOON states, path entangled states of
the form $\frac{1}{\sqrt{2}}\left( \left| N\right\rangle
_{k1}\left| 0\right\rangle _{k2}+\left| 0\right\rangle _{k1}\left|
N\right\rangle _{k2}\right) $, of fundamental relevance since a
single-photon phase shift $\varphi $ induces a relative shift
between the two components equal to $N\varphi $. This feature can
be exploited to enhance phase resolution in interferometric
measurements, leading to a sub-Rayleigh resolution scaling as
$\frac{\lambda }{2N}$; $\lambda $ being the wavelength of the
field \cite{DAng01}. The generation of photonic NOON states has
been the subject of intense theoretical research \cite{Kapa06},
but up to now the actual experimental implementation has been
limited to a posteriori generation of two, three and four photons
states \cite {Edam02,Mitc04,Walt04} and to the conditional
generation of a NOON state with $N=2$ \cite{Eise05}. The weak
value of the generated number of photons strongly limits the
potential applications to quantum lithography and quantum
metrology. As alternative approach to emulate the quantum
resolution, it has been proposed to adopt classical, coherent
light \cite {Bent04} effectively exploiting the non-linearity of
the absorbing material. First experimental results have been
recently reported: Yablonovitch et al. proposed to use an
interference technique with multiple-frequency beams achieving an
experimental visibility of $3\%$ \cite{Koro04} while Boyd et al.
achieved a two-fold enhancement of the resolution over the
standard Rayleigh limit adopting a UV lithographic material
excited by multi-photon aborption \cite{Boyd06}.

\begin{figure}[h]
\includegraphics[scale=.3]{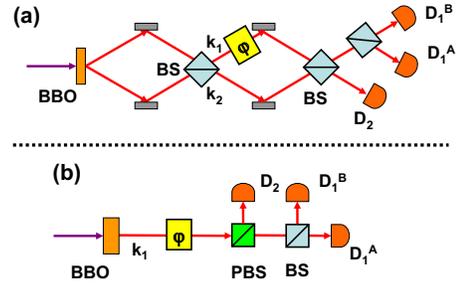}
\caption{ (a) Experimental scheme for quantum lithography based on
spontaneous parametric down-conversion. The 2-photon absorption is simulated
through a 2-photon coincidence detection. (b) Configuration based on
polarization entangled beams. }
\end{figure}

Recently it has been proposed to exploit a high gain optical parametric
amplifier acting on the vacuum field to generate fields with a high number
of photons still exhibiting sub-Rayleigh periods: Fig.1-a \cite{Agar01}. The
quantum lithography architecture involves the generation of correlated beams
over the modes $k_{1}$ and $k_{2}$, the mixing over a beamsplitter (BS), the
phase shifting of one mode and then the recombination of the two modes over
a second \textrm{BS}. Finally the output state is detected via a two-photon
absorption over the outgoing modes. In the typical low-gain regime the
quantum state $2^{-1/2}\left( \left| 2\right\rangle _{k1}\left|
0\right\rangle _{k2}-\left| 0\right\rangle _{k1}\left| 2\right\rangle
_{k2}\right) $ is generated by spontaneous parametric down-conversion (SPDC)
and a Hong-Ou-Mandel interferometer adopting a post-selection technique \cite
{Edam02}. In Ref. \cite{Agar01,Naga01,Agar06} it has recently been
theoretically shown that, for any gain of the parametric amplifier, the
two-photon excitation rate presents a fringe pattern of the form $1+\cos
2\varphi $ which never falls below a visibility of $20\%$. Such an approach
could lead to high resolution with a high number of photons overcoming the
difficulty connected with the adoption of NOON state with a low number of
photons. At variance with the scheme based on classical radiation, the one
here demonstrated exploits quantum features of the OPA field.

\section{Sub-Rayleigh resolution by an unseeded high-gain optical parametric
amplifier}

In the present paper we experimentally investigate the properties of the
optical parametric amplifier (OPA) operating in the high gain regime.
Instead of dealing with path entangled states, we consider the generation of
entangled states over the same mode but with orthogonal polarizations,
respectively, horizontal ($\overrightarrow{\pi }_{H}$) and vertical ($%
\overrightarrow{\pi }_{V}$). The scheme is shown in Fig.1-b. The non-linear
crystal (BBO) is pumped by a high-power laser. The first order contribution
to the output field is the twin photons state over the same mode: $%
2^{-1/2}\left( \left| 2\right\rangle _{+}\left| 0\right\rangle _{-}+\left|
0\right\rangle _{+}\left| 2\right\rangle _{-}\right) =\left| 1\right\rangle
_{H}\left| 1\right\rangle _{V}$, with $\overrightarrow{\pi }_{\pm }=2^{-1/2}(%
\overrightarrow{\pi }_{H}\pm \overrightarrow{\pi }_{V})$. Such
polarization-entangled photons can be easily converted to path entangled via
an additional polarizing \textrm{BS}. The \textrm{PBS }of Fig1.b mixes the
two polarization components $\overrightarrow{\pi }_{+}$ and $\overrightarrow{%
\pi }_{-}$ as the second \textrm{BS} of Fig1.a mixes the two different path
modes, similar arrangements of polarization NOON were adopted in\ Ref.\cite
{Mitc04,Walt04}.

\begin{figure}[t]
\includegraphics[scale=.3]{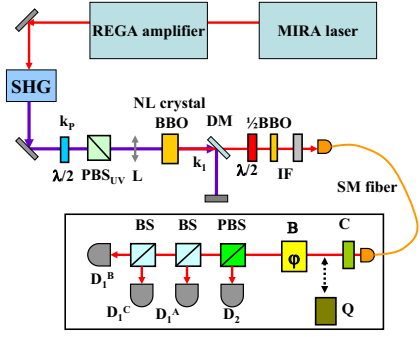}
\caption{Experimental layout. The radiation generated in the NL crystal is
spectrally filtered by an interferential filter (IF) with bandwidth equal to
3nm and spatially selected adopting a single mode (SM) fiber.}
\end{figure}

\begin{figure}[b]
\includegraphics[scale=.28]{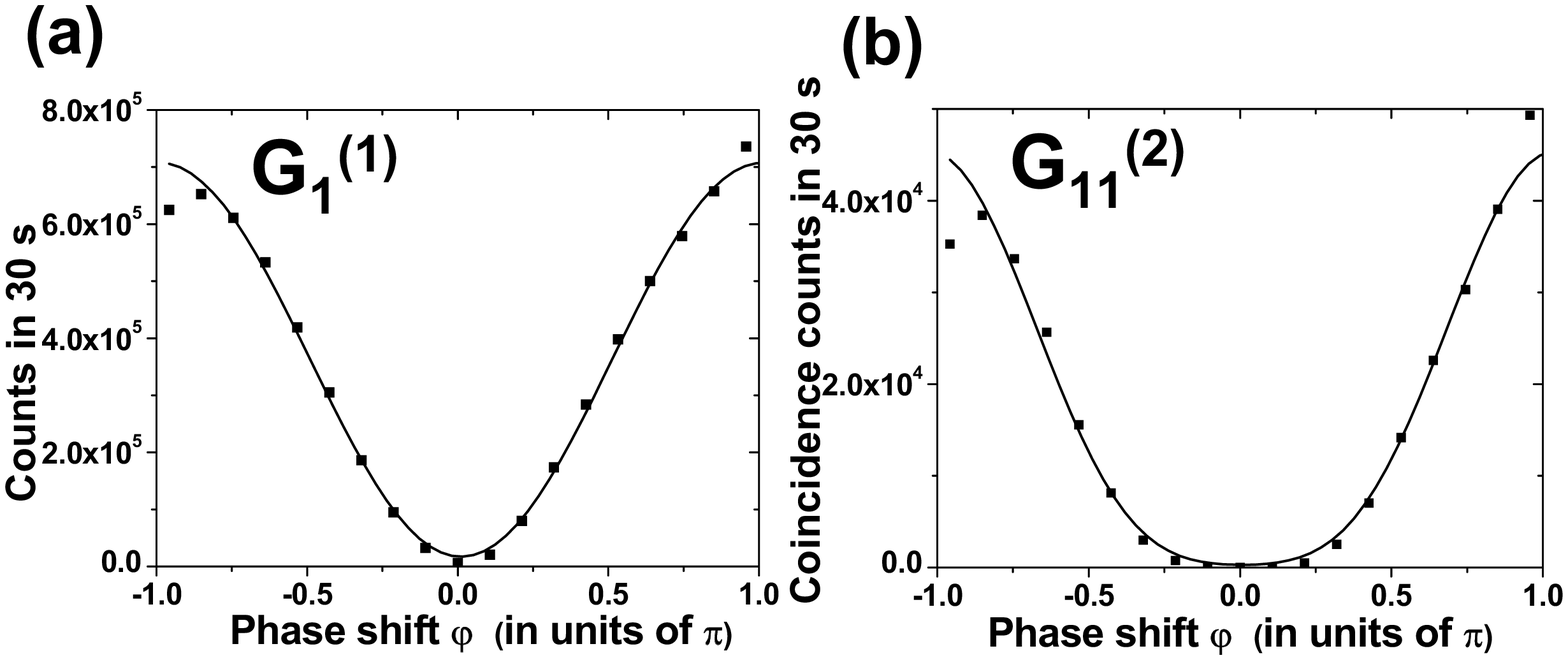}
\caption{Coherent state as input field. (a) $G_{1}^{(1)}$: Count rates $%
D_{1}^{A}$ versus the phase $\protect\varphi $. (b) $G_{11}^{(2)}$:
Coincidence counts $[D_{1}^{A},D_{1}^{B}]$ versus the phase $\protect\varphi %
.$}
\end{figure}

Let us introduce the generic quantum states generated by the OPA acting on
the vacuum fields $\left| 0\right\rangle _{H}\left| 0\right\rangle _{V}$.
The interaction Hamiltonian of the optical parametric amplification $%
\widehat{H}_{coll}=i\chi \hbar \widehat{a}_{H}^{\dagger }\widehat{a}%
_{V}^{\dagger }+h.c.$ acts on the single spatial mode $\mathbf{k}_{1}.$ A
fundamental physical property of $\widehat{H}_{coll}$ consists of its
expression for any polarization basis belonging to the equatorial basis.
Indeed $\widehat{H}_{coll}$ can be written as $\frac{1}{2}i\chi \hbar \left(
\widehat{a}_{+}^{\dagger 2}+\widehat{a}_{-}^{\dagger 2}\right) +h.c.$ where $%
\widehat{a}_{\pm }^{\dagger }$ are the creation operators for the $%
\overrightarrow{\pi }_{\pm }$ polarization modes, respectively. The output
state over the mode $\mathbf{k}_{1}$ of the unseeded optical parametric
amplifier is found to be:
\begin{equation}
\left| \Phi \right\rangle =\frac{1}{C}\sum_{n=0}^{+\infty }\Gamma ^{n}\left|
n\right\rangle _{H}\left| n\right\rangle _{V}=\frac{1}{C}\sum_{i,j=0}^{+%
\infty }\Gamma ^{i+j}\left| 2i\right\rangle _{+}\left| 2j\right\rangle _{-}
\end{equation}
with $C\equiv \cosh g$, $\Gamma \equiv \tanh g$, being $g$ the non linear
(NL) gain \cite{DeMa05}. This state is usually called a two-mode squeezed
state. The average photon number created per polarization mode is equal to $%
\overline{n}=\sinh ^{2}g$. In the interferometric setup the output state is
shifted by a phase $\varphi $ in the basis $\{\overrightarrow{\pi }_{+},%
\overrightarrow{\pi }_{-}\}$. Hence the output state is detected in the
basis $\{\overrightarrow{\pi }_{H},\overrightarrow{\pi }_{V}\};$ adopting a
polarizing beam splitter (PBS).\ For different values of the phase $\varphi $%
, the quantum state $\left| \Phi \right\rangle $ is analyzed through two
different second-order correlation functions $G_{12}^{(2)}=\left\langle \Phi
\right| c_{1}^{\dagger }c_{2}^{\dagger }c_{2}c_{1}\left| \Phi \right\rangle $
and $G_{11}^{(2)}=\left\langle \Phi \right| c_{1}^{\dagger }c_{1}^{\dagger
}c_{1}c_{1}\left| \Phi \right\rangle $ where $\{c_{1}^{\dagger }=\left( \cos
\frac{\varphi }{2}\widehat{a}_{H}^{\dagger }-i\sin \frac{\varphi }{2}%
\widehat{a}_{V}^{\dagger }\right) e^{i\frac{\varphi }{2}},c_{2}^{\dagger
}=\left( -i\sin \frac{\varphi }{2}\widehat{a}_{H}^{\dagger }+\cos \frac{%
\varphi }{2}\widehat{a}_{V}^{\dagger }\right) e^{i\frac{\varphi }{2}}\}$ are
the output modes of the PBS. By tuning $\varphi $ it is found $%
G_{12}^{(2)}(\varphi )=\overline{n}^{2}+\frac{1}{2}(\overline{n}^{2}+%
\overline{n})(1+\cos 2\varphi )$ and $G_{11}^{(2)}(\varphi )=2\overline{n}%
^{2}+\frac{1}{2}(\overline{n}^{2}+\overline{n})(1-\cos 2\varphi )$ \cite
{Naga01}. The corresponding visibilities of the obtained fringe patterns
read $V_{1}^{(2)}=\frac{\overline{n}+1}{5\overline{n}+1}$ and $V_{12}^{(2)}=%
\frac{\overline{n}+1}{3\overline{n}+1}$. We observe that a non-vanishing
visibility is found for any value of $g:$ $V_{1}^{(2)}(g\rightarrow \infty )=%
\frac{1}{5}$ and $V_{12}^{(2)}(g\rightarrow \infty )=\frac{1}{3}$. The two
fringe patterns exhibit a dependence on $2\varphi $ and hence a period equal
to $\frac{\lambda }{2}$. This feature can be exploited to carry out
interferometry with sub-Rayleigh resolution, i.e., with fringe period lower
than $\lambda $, in a higher flux regime compared to the two photon
configurations.

\begin{figure}[b]
\includegraphics[scale=.32]{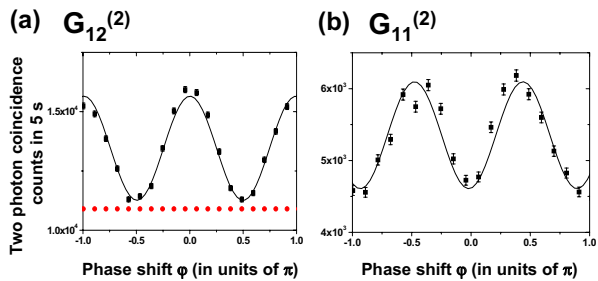}
\caption{ (a) $G_{12}^{(2)}$:\ two-photon coincidence counts $%
[D_{1}^{A},D_{2}]$ versus the phase $\protect\varphi $ introduced by the
Soleil-Babinet compensator $(g=1.4)$. Circle data: expected accidental
(without correlations) coincidence rates. (b) $G_{11}^{(2)}:$ two-photon
coincidence counts $[D_{1}^{A},D_{1}^{B}]$ versus the phase $\protect\varphi
$ $(g=1.4)$. Variations of the detectors signal $(\sim 10\%)$ due to
different couplings of polarizations $\vec{\protect\pi}_{H}$ and $\vec{%
\protect\pi}_{V}$ with fiber have been corrected by normalizing coincidence
counts with signal rates.}
\end{figure}

\section{Experimental setup and results}

We now briefly describe the experimental configuration: Fig.2. The
excitation source was a Ti:Sa Coherent MIRA mode-locked laser further
amplified by a Ti:Sa regenerative REGA device operating with pulse duration $%
180$\textrm{fs} at a repetition rate of $250$\textrm{kHz}. The output beam,
frequency-doubled by second harmonic generation, provided the excitation
beam of UV wavelength $\lambda _{P}=397.5$\textrm{nm} and power $300$\textrm{%
mW}. The horizontally polarized UV beam was then adopted to pump a
non-linear BBO crystal, which generated pairs of photons with polarization $%
\vec{\pi}_{H}$ and $\vec{\pi}_{V}$ over the modes $k_{1}$ with same
wavelength $\lambda =2\lambda _{P}$. This source allows us to obtain a high
value of the gain $g,$ which depends on the pumping power: $g\varpropto
\sqrt{P_{UV}}.$ Compared to conventional pulsed sources used to pump SPDC
process, which achieve $g\simeq 0.1$, the energy per pulse is enhanced by a
factor $\simeq 400$, leading to an increase of the gain value in the range
of $20-40$ depending on the focal length of the UV beam. The pumping power
could be tuned adopting a half-wave plate and a polarizing beam splitter (%
\textrm{PBS}$_{UV}$). The output state of BBO crystal with wavelength $%
\lambda $ was spatially separated by the fundamental UV\ beam through a
dichroic mirror (\textrm{DM}), then spectrally filtered adopting an
interferential filter (\textrm{IF}) with bandwidth equal to $3$\textrm{nm}
and then coupled to a single mode fiber in order to select spatially a
single mode of emission. A $\lambda /2$ waveplate and a BBO with thickness
of $0.75$\textrm{mm} provided the compensation of walk-off effects. At the
output of the fiber, after compensation (\textrm{C}) of the polarization
rotation induced by the fiber, a phase shifting $\varphi $ was introduced
adopting a Soleil-Babinet compensator (\textrm{B}). The output radiation was
then analyzed through a polarizing beam splitter (\textrm{PBS}) and detected
adopting single photon detectors SPCM-AQR14 ($\mathrm{D}_{1}^{A}\mathrm{,D}%
_{1}^{B}\mathrm{,D}_{1}^{C}\mathrm{,D}_{2}$). To characterize the detection
apparatus, a coherent state with wavelength $\lambda $ and polarization $%
\vec{\pi}_{H}$ was fed into the mode $k_{1}.$ The count rates $D_{1}^{A}$
and the coincidence rates $[D_{1}^{A},D_{1}^{B}]$ were measured versus the
phase $\varphi $: Fig.3. High visibility patterns have been observed with a
period equal to $\lambda $.

\begin{figure}[h]
\includegraphics[scale=.32]{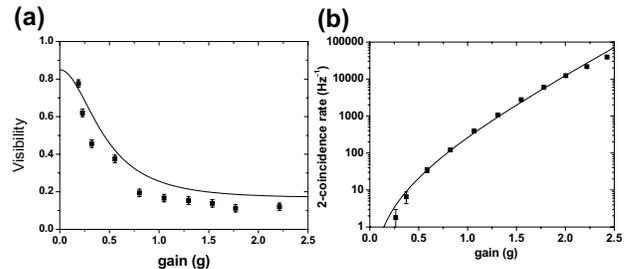} \vspace{0.5cm}
\caption{ (a) Visibility versus non-linear gain. The continuous line
corresponds to the function $V_{\max }\times V_{1}^{(2)}(g).$ (b) Excitation
rate versus non-linear gain. The continuous line corresponds to the function
$\overline{R}^{(2)}(\protect\alpha g)$ where the parameter $\protect\alpha $
has been optimized by fitting the data and reads $0.85.$ }
\end{figure}

As a first experimental step we have characterized the two-photon state
generated by SPDC in the low gain regime. The visibilities have been found $%
V_{12}^{(2)}=(85\pm 2)\%$ and $V_{1}^{(2)}=(80\pm 1)\%$. The discrepancy
with the expected values $V=1$ is attributed to double-pair emission and to
experimental imperfections. The same measurement has been carried out
increasing the UV\ pump beam in order to measure the fringe patterns for
different gain values. The gain has been estimated with the method
introduced in Ref. \cite{Eise04,Cami06}.\ Fig. 4 refers to the configuration
$g=1.4$. The visibilities have been found to be $V_{12}^{(2)}=(16.8\pm
0.6)\% $ and $V_{1}^{(2)}=(15\pm 1)\%$. The sub Rayleigh resolution is
clearly shown by the experimental data of Fig.4. For the sake of
completeness\ the value of $V_{1}^{(2)}$ has been measured for different
gains: Fig.5-a. The continuous line shows the expected theoretical function $%
V_{1}^{(2)}(g)$ multiplied by the extrapolated visibilities for $%
g\rightarrow 0:$ $V_{\max }=0.85$. We attribute the discrepancy between
experimental and theoretical visibilities to partial multimode operation of
the optical parametric amplifier \cite{Than04}.

\begin{figure}[h]
\includegraphics[scale=.28]{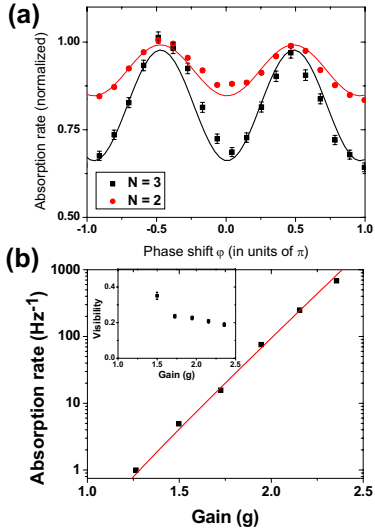}
\caption{(a) Circle data: $G_{1}^{(2)}$\ (two-photon coincidences $%
[D_{1}^{A},D_{1}^{B}]$) versus the phase $\protect\varphi $; Square data: $%
G_{1}^{(3)}$\ (three-photon coincidences $[D_{1}^{A},D_{1}^{B},D_{1}^{C}]$)
versus the phase $\protect\varphi $ $(g=2.4)$ (b) Excitation rate versus
non-linear gain. The continuous line corresponds to the function $\overline{R%
}^{(3)}(g)$. Inset: visibility $V_{1}^{(3)}$ versus non-linear gain. }
\end{figure}

To estimate the dependence of the excitation rate from the NL gain $g$, the
maxima and the minima of the fringes have been measured versus the pumping
power. The average data are reported in Fig.5-b. Our experiment validates
the result found by Agarwal et al. \cite{Agar06} showing an exponential
dependence on the parameter $g$. The expected two-photon excitation rate
reads $\overline{R}^{(2)}=2\sigma ^{(2)}(\overline{n}+5\overline{n}^{2})$
where $\sigma ^{(2)}$ is a generalized two-photon excitation cross section
\cite{Agar06}. Fig.5-b shows how the excitation efficiency scales
quadratically with the light intensity, in contrast with the two-photon SPDC
regime; which leads to a linear dependence \cite{Java90,Peri98,Daya05}.

As further demonstration of the potentialities of the present approach, the
simultaneous detection of three photons over the same mode has been
investigated. The average three photon excitation rate reads $\overline{R}%
^{(3)}=12\sigma ^{(3)}(7\overline{n}^{3}+3\overline{n}^{2})$ where $\sigma
^{(3)}$ is a generalized three-photon excitation cross section and the
visibility of the fringes is theoretically found as $V_{1}^{(3)}=\frac{3%
\overline{n}+3}{7\overline{n}+3}$ \cite{Agar06}$.$ Again a non-vanishing
value of $V_{1}^{(3)}$ is found for any value of $g:$ $V_{1}^{(3)}(g%
\rightarrow \infty )=\frac{3}{7}$ and the patterns exhibit a period equal to
$\frac{\lambda }{2}$. Furthermore an increase of visibility is expected $%
V_{1}^{(3)}>V_{1}^{(2)}$. To demonstrate such a feature the three-photon
coincidence rate $G_{1}^{(3)}$ has been measured versus the phase $\varphi $%
: Fig.6-a$.$ An increase of the visibility has been found $%
V_{1}^{(3)}=(21.6\pm 0.6)\%$, the experimental dependencies of the three
photon absorption rates and visibilities are shown in Fig.6-b.

In order to investigate the connection between the quantum feature of the
state and the visibility of the fringe pattern, we have introduced a
decoherence between the two polarization components $\{\vec{\pi}_{H},\vec{\pi%
}_{V}\}$ or $\{\vec{\pi}_{+},\vec{\pi}_{-}\}$ with a quartz crystal (\textrm{%
Q}) with a length equal to $20$\textrm{mm}$.$ This device introduces a
temporal delay higher than the coherence time of the multiphoton fields. The
$G_{1}^{(2)}$ has been first measured without the quartz: Fig.7. When the
decoherence affects the $\{\vec{\pi}_{H},\vec{\pi}_{V}\}$ components, we
observe a reduction of visibility down to $(4.8\pm 0.6)\%$; on the other
hand when the decoherence involves the $\{\vec{\pi}_{+},\vec{\pi}_{-}\}$ we
observe the disappearance of the fringe patterns. The inversion of the
maxima and minima of the square data compared to the circle data, a
phenomenon not expected in the single-pair regime, is due to the reduction
of the bunching effect among photons detected on the same polarization mode
when decoherence is introduced.

\begin{figure}[h]
\includegraphics[scale=.2]{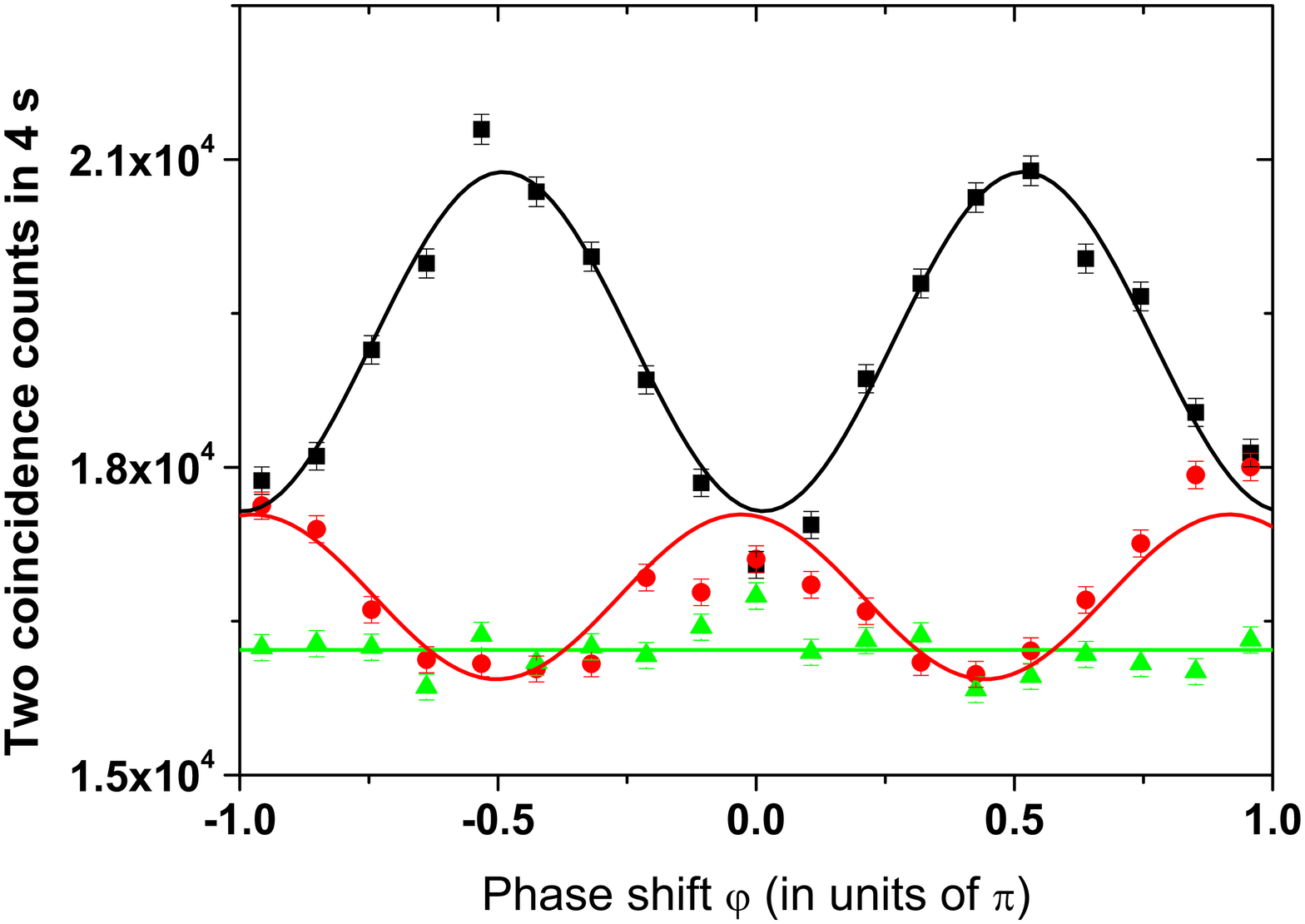}
\caption{Effects of coherence on fringe visibility. Square data: 2-photon
coincidences counts versus the phase $\protect\varphi $ without introducing
decoherence. Circle data: decoherence introduced between the polarization
components$\{\vec{\protect\pi }_{H},\vec{\protect\pi }_{V}\}$. Triangle
data: decoherence introduced between $\{\vec{\protect\pi }_{+},\vec{\protect%
\pi }_{-}\}$.}
\end{figure}

\section{Conclusions and perspectives}

In the present paper we have experimentally demonstrated that the output
state of a high-gain optical parametric amplifier can be intense; yet
exhibits quantum features. We have extended the previous implementations
\cite{Edam02}, limited to the a-posteriori generation of a two photon state
to a large number of photons. The reduced visibility has been increased by
means of three-photon absorption as suggested by \cite{Agar06}. In this
situation a higher visibility has been achieved but without an increase of
the fringe density. The present article opens interesting perspectives for
the implementation of a quantum interferometric setup. By adopting homodyne
detection on the output fields, the Heisenberg limit for the phase
resolution can be achieved \cite{Steu04}. Recently it has been proposed to
combine high-gain SPDC\ with coherent states to create general NOON states
with a fidelity of about $94\%$ \cite{Hofm07}. This work was supported by
the PRIN 2005 of MIUR and Progetto Innesco 2006 (CNISM),\ ARO; DTO\ and
DARPA.

\end{document}